\documentstyle[aps,prl,twocolumn,graphicx,floats,epsfig,times]{revtex}

\begin{document}

\draft 

\onecolumn 

\noindent

\title{Observation of Direct Photons in Central 158$~${\it A}~GeV
  $^{208}$Pb\/+\/$^{208}$Pb Collisions}

\author{ M.M.~Aggarwal,$^{1}$ A.~Agnihotri,$^{2}$ Z.~Ahammed,$^{3}$
  A.L.S.~Angelis,$^{4}$ V.~Antonenko,$^{5}$ V.~Arefiev,$^{6}$
  V.~Astakhov,$^{6}$ V.~Avdeitchikov,$^{6}$ T.C.~Awes,$^{7}$
  P.V.K.S.~Baba,$^{8}$
% ...
}

\author{(WA98 Collaboration)}

\address{$^{1}$~University of Panjab, Chandigarh 160014, India}
\address{$^{2}$~University of Rajasthan, Jaipur 302004, Rajasthan,
  India} \address{$^{3}$~Variable Energy Cyclotron Centre, Calcutta
  700 064, India} \address{$^{4}$~University of Geneva, CH-1211 Geneva
  4,Switzerland} \address{$^{5}$~RRC ``Kurchatov Institute'', RU-123182 Moscow,
  Russia} \address{$^{6}$~Joint Institute for Nuclear Research,
  RU-141980 Dubna, Russia} \address{$^{7}$~Oak Ridge National
  Laboratory, Oak Ridge, Tennessee 37831-6372, USA}
\address{$^{8}$~University of Jammu, Jammu 180001, India}
% ...

\date{Draft 1.0, \today} \maketitle
\begin{abstract}
  A measurement of direct photon production in
  $^{208}$Pb\/+\/$^{208}$Pb collisions at 158~{\it A}~GeV has been
  carried out in the CERN WA98 experiment.  The invariant yield 
  of direct photons in central collisions is extracted
  as a function of transverse momentum
  in the interval $0.5 < p_T < 4$ GeV/c.  A significant direct photon
signal, compared to statistical and systematical errors, is seen
at  $ p_T > 1.5$ GeV/c. The results constitute the
  first observation of direct photons in ultrarelativistic 
  heavy-ion collisions which could be significant for diagnosis
of quark gluon plasma formation.
\end{abstract}
\twocolumn

The observation of a new phase of strongly interacting matter, the 
quark gluon plasma (QGP), is one 
of the most important goals of current nuclear physics research. 
An extensive experimental program has been undertaken at the 
CERN SPS accelerator with Pb-ion beams of 158$A$GeV to search
for and investigate the QGP. 
Several observations, such as
suppression of the J/$\psi$ resonance~\cite{Abr97} and the 
enhancement of strangeness~\cite{And99}, hint at an interesting 
new behavior of the matter produced in these collisions.
While such observations imply a hot and dense 
initial phase with strong rescattering, consistent
with the assumption that a quark gluon plasma 
was formed, a direct signature of the plasma and its properties is
missing. It is therefore of great interest to search for 
photons emitted directly from the early hot phase of the relativistic 
heavy-ion collisions.

Photons (both real and virtual) were one of the earliest proposed 
signatures for QGP formation \cite{Fei76,Shu78}. Real photons are 
dominantly produced by scatterings of charged particles during the 
collision. Once produced, they interact with the surrounding matter 
only electromagnetically resulting in a long mean free path. They 
are therefore likely to escape from the system directly after 
production without further interaction, unlike hadrons. 
Thus, photons carry information on their emitting sources from 
throughout the entire collision history, including the initial hot 
and dense phase.

Following early estimates of photon emission 
rates~\cite{Kaj81,Hal82,Sin83,McL85}, Kapusta et al. 
\cite{Kap91} made detailed comparisons of the emissivity of the 
QGP and a hadron gas as two contrasting scenarios. It was demonstrated 
that the thermal emission rates of a hadron gas and a QGP 
were very similar and dependent essentially only on  
the temperature $T$. This led the authors to conclude that 
direct photons 
are a good thermometer for strongly interacting matter, but
would not in themselves allow to distinguish between the two scenarios.
%

Recently, it was shown by Aurenche et al.~\cite{Aur98} that 
photon production rates in the QGP when calculated up to two loop
diagrams, 
are considerably greater than the earlier lowest order
estimates. A new higher order process of $q\overline{q}$ annihilation
with rescattering was found to dominate the photon emission rate
from quark matter at high photon energies. Following this result,  
Srivastava~\cite{Sri99} has reinvestigated the predicted
photon production in heavy-ion collisions and shown that at
sufficiently high initial temperatures
the photon yield from quark matter may significantly exceed
the contribution from the hadronic matter to provide a direct probe of
the quark matter phase.

A large number of measurements of prompt photon production at 
high transverse momentum ($p_{T} > 3 \, \mathrm{GeV}/c$) exist
for proton-proton, proton-antiproton, and proton-nucleus collisions
(see e.g. \cite{VW}). To a great extent, especially at 
higher $\sqrt{s}$, these data 
can be successfully described by perturbative QCD calculations
and provide an important foundation from which to study photon
production in nucleus-nucleus collisions.
First attempts to observe direct photon production 
in ultrarelativistic heavy-ion collisions with oxygen and sulphur
beams found 
no significant excess~\cite{Ake90,Alb91,Bau96,Alb96}.
The WA80 collaboration \cite{Alb96} provided the most interesting
result with a $p_{T}$ dependent upper limit on the direct
photon production in S+Au collisions at 200$A$GeV. 
This result was subsequently used by several authors to rule out a
simple version of the hadron gas scenario 
\cite{ref:sr94,ref:ar95,ref:du95,ref:ne95} and has been interpreted 
to set an upper limit on the initial temperature of $T_{i} = 250 
\,\mathrm{MeV}$ \cite{ref:so97}. 

In this paper we report on the first observation of direct 
photon production in ultrarelativistic heavy-ion collisions.
The results are from the CERN experiment WA98 \cite{misc:wa98:proposal:91} 
which consists
of large acceptance photon and hadron spectrometers. In addition, several 
other large acceptance devices allow to measure various global
variables on an event-by-event basis for event characterization. 
Photons are measured with the WA98 lead-glass photon detector,
LEDA, which consisted of 10,080 individual modules with 
photomultiplier readout. The detector was located at a distance of 
21.5~m from the target and covered the pseudorapidity interval 
$2.35 < \eta < 2.95$ $(y_{cm}=2.9)$. The particle identification was 
supplemented by a charged particle veto detector in front of LEDA.

The results presented here were obtained from an analysis of the
data taken with Pb beams in 1995 and 1996.
The 20\% most peripheral and the 10\% most central reactions have been 
selected from the minimum bias cross section 
($\sigma_{min.bias} \approx 6300 \, \mathrm{mb}$) 
using the measured transverse energy $E_{T}$. 
In total, $\approx 6.7 \cdot 10^{6}$ central and $\approx 4.3 \cdot 
10^{6}$ peripheral reactions have been analyzed.

The extraction of direct photons in the high multiplicity environment 
of heavy-ion collisions must 
be performed on a statistical basis by comparison of the measured
inclusive photon
spectra to the background expected from hadronic decays. 
Individual photons cannot 
be tagged as isolated direct photons in these reactions due to
the high multiplicities. 
To obtain the direct photon spectrum the 
following steps are performed (for a detailed description of the 
detectors and the analysis procedure see \cite{wa98photonslong}):
First, the raw photon spectra are accumulated after application 
of the photon identification criteria (such as transverse shower size)
to the showers observed in the LEDA. 
The raw photon spectra are then corrected for contamination by 
charged and neutral hadrons, for conversions, for the 
identification efficiency, and acceptance. 
The efficiency includes all effects of the detector 
response such as distortions by shower overlap, dead and bad 
modules, and energy resolution.
Neutral pions are reconstructed via their $\gamma\gamma$ decay branch. 
Invariant mass spectra are accumulated for all photon pairs for each
pair $p_T$ bin.
The photon-pair combinatorial background is estimated by event-mixing and then 
subtracted from the real-pair spectra. The yield in the  $\pi^0$
mass peak is extracted 
to obtain the raw neutral pion $p_T$ spectra. These are then 
corrected for conversions, for the $\pi^0$ identification efficiency, and 
for geometrical acceptance. In addition, $\eta$ mesons are extracted in a 
limited transverse momentum range with an analogous procedure. 

\begin{figure}[bt]
\begin{center}
  \includegraphics[scale=0.4]{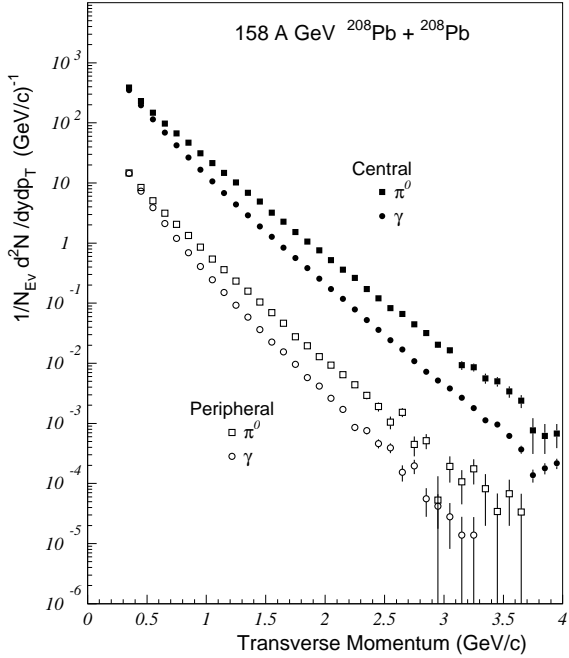}
\vspace{0.3cm}
  \caption{The inclusive photon (circles) and $\pi^0$ (squares) 
  transverse momentum distributions 
  for peripheral (open points) and central (solid points)
  158$~${\it A}~GeV $^{208}$Pb\/+\/$^{208}$Pb collisions. 
  The data have been corrected
  for efficiency and acceptance. Only statistical errors are shown.
  }
\label{fig:photon_pi0_pt}
\end{center}
\end{figure}

The final measured inclusive photon spectra are 
then compared to the calculated background photon 
spectra to check for a possible photon excess beyond that from 
long-lived radiative decays.
The background calculation is based on the 
measured  $\pi^0$ spectra and the measured $\eta/\pi^0$-ratio. 
The spectral shapes of other hadrons having radiative decays are calculated
assuming $m_{T}$-scaling~\cite{Bor76} with yields relative to
$\pi^0$'s taken from the literature.
It should be noted that the measured 
contribution (from $\pi^0$ and $\eta$) amounts to $\approx 97 \% $ of 
the total photon background. 

\begin{figure}[hbt]
\begin{center}
  \includegraphics[scale=0.4]{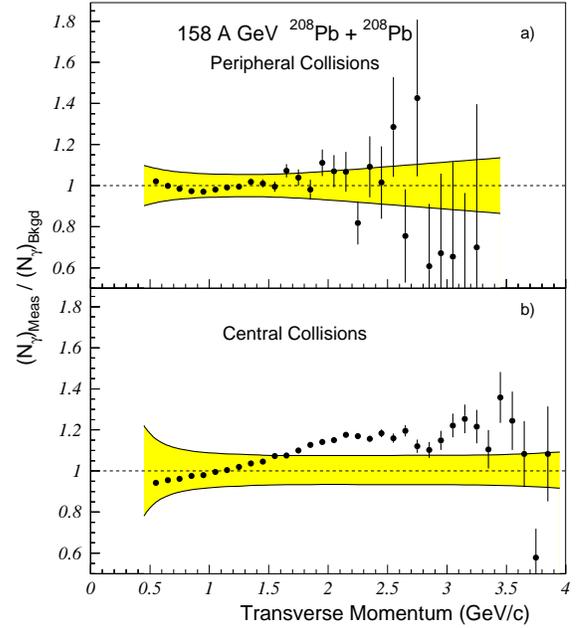}
\vspace{0.3cm}
\caption{The $\gamma_{\rm Meas}/\gamma_{\rm Bkgd}$ ratio 
  as a function of transverse momentum for peripheral (part a)) and
  central (part b)) 158~{\it A}~GeV $^{208}$Pb\/+\/$^{208}$Pb collisions. The
  errors on the data points indicate the statistical errors only. The
  $p_T$-dependent systematical errors are indicated by the shaded bands.
  }
\label{fig:gamma_excess}
\end{center}
\end{figure}

Fig.~\ref{fig:photon_pi0_pt} shows the fully corrected inclusive photon 
spectra for peripheral and central collisions. The spectra cover the 
$p_{T}$ range of $0.3 - 4.0 \,\mathrm{GeV}/c$ (slightly less for 
peripheral collisions) and extend over six orders of magnitude. 
Fig.~\ref{fig:photon_pi0_pt} also shows the distributions of neutral 
pions which extend over a similar momentum range with slightly 
larger statistical errors. 

The ratio of measured photons to calculated background photons is 
displayed in Fig.~\ref{fig:gamma_excess} as a function of transverse 
momentum. The upper plot shows the ratio for peripheral collisions 
which is seen to be compatible with one, i.e. no indication of a 
direct photon excess is observed. The lower plot shows the same ratio for 
central collisions. It rises from a value of $\approx 1$ at low 
$p_{T}$ to exhibit an excess of about 20\% at high $p_{T}$.

A careful study of possible systematical errors is crucial for
the direct photon analysis. The various sources of systematical 
errors have been 
investigated and are summarized in Table~\ref{table3}. The largest 
contributions are from the $\gamma$ and $\pi^0$ identification 
efficiencies and the uncertainties related to the $\eta$ measurement. 
It should be emphasized
that the inclusive photon and neutral meson
(the basis for the background calculation) yields have been extracted from 
the same detector for exactly the same data sample. 
This decreases the sensitivity to many detector related errors and 
eliminates all errors associated  
with trigger bias or absolute yield normalization. The estimate of the
systematical errors has been checked by performing the entire 
analysis with various photon selection criteria which change the
efficiency and background corrections by factors of 2-3. The final 
results were verified to be consistent within the 
systematical errors for the different 
analysis cuts.  Full details on 
the systematical error estimates are given in \cite{wa98photonslong}. 
The total $p_{T}$-dependent systematical errors are shown by the shaded 
regions in Fig.~\ref{fig:gamma_excess}. A significant photon 
excess is clearly observed in central collisions for $p_{T} > 1.5 \, 
\mathrm{GeV}/c$.

\begin{figure}[hbt]
\begin{center}
  \includegraphics[scale=0.4]{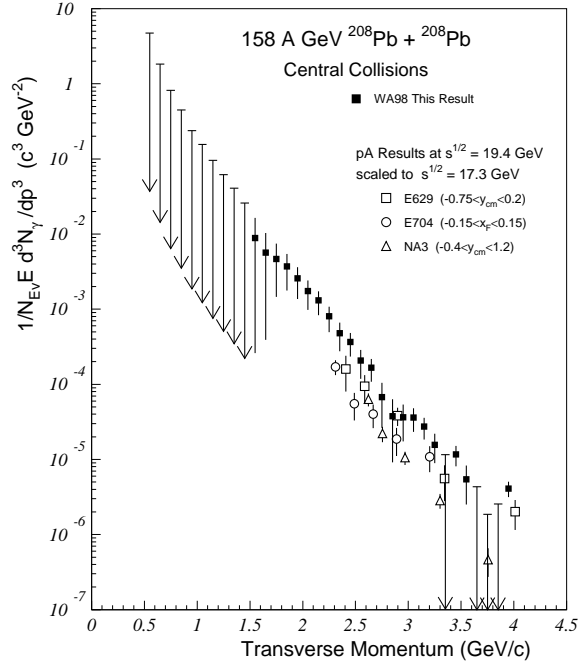}
\vspace{0.3cm}
\caption{The invariant direct photon multiplicity 
  for central 158~{\it A}~GeV $^{208}$Pb\/+\/$^{208}$Pb collisions.
  The error bars indicate the combined statistical and systematical
  errors.  Data points with downward arrows indicate unbounded 90\%
  CL upper limits.  Results of several direct photon measurements for
  proton-induced reactions have been scaled to central
  $^{208}$Pb\/+\/$^{208}$Pb collisions for comparison. 
  }
\label{fig:gamma_excess_cs}
\end{center}
\end{figure}

The final  invariant direct photon yield per central 
collision is presented in Fig.~\ref{fig:gamma_excess_cs}. 
The statistical and asymmetric systematical
errors of Fig.~\ref{fig:gamma_excess} are added in quadrature to
obtain the total upper and lower errors shown in
Fig.~\ref{fig:gamma_excess_cs}. An additional $p_T$-dependent error is
included to account for that portion of the uncertainty in the energy scale
which cancels in the ratios. In the case that the
lower error is less than zero a downward arrow is shown with the tail
of the arrow indicating the 90\% confidence level upper limit
($\gamma_{Excess}+1.28\,\sigma_{Upper}$).

No published prompt photon results exist for proton-induced reactions
at the $\sqrt{s}$ of the present measurement.
Instead, prompt photon yields for proton-induced reactions on fixed
targets at 200 GeV are shown in Fig.~\ref{fig:gamma_excess_cs} for
comparison. 
Results are shown from FNAL experiment E704~\cite{plb:ada95}
for proton-proton reactions, and from FNAL experiment
E629~\cite{prl:mcl83}  and CERN SPS experiment 
NA3~\cite{zpc:bad86} for proton-carbon reactions.
These results have been divided by the total pp inelastic cross
section ($\sigma_{int}=30$~mb) and by the mass number of the 
target to obtain the invariant direct photon yield per nucleon-nucleon
collision. They have then been multiplied by the calculated average number 
of nucleon-nucleon collisions (660) for the central Pb+Pb event selection
for comparison with the present measurements. This scaling is
estimated to have an uncertainty of less than 10\%.
The proton-induced results have also been scaled 
from $\sqrt{s}=19.4$ GeV to the lower $\sqrt{s}=17.3$ GeV of the 
present measurement under the assumption that $E
d^3\sigma_{\gamma}/dp^3 = f(x_T)/s^2$, where  
$x_T=2p_T/\sqrt{s}$\cite{rmp:owe87}. The
$\sqrt{s}$-scaling effectively reduces the $19.4$ GeV 
proton-induced results by about a factor of two. 
This comparison indicates that the observed direct photon
production in central $^{208}$Pb\/+\/$^{208}$Pb collisions
has a shape similar to that expected for proton-induced reactions
at the same $\sqrt{s}$ but a yield which is enhanced.

In summary, the first observation of direct photons 
in ultrarelativistic heavy-ion collisions has been presented. 
While peripheral Pb+Pb 
collisions exhibit no significant photon excess, the 10\% most central 
reactions show a clear excess of direct photons in the range of 
$p_T$ greater than about $1.5 \, \mathrm{GeV}/c$. The invariant
direct photon multiplicity as a function of transverse momentum
was presented for central $^{208}$Pb\/+\/$^{208}$Pb 
collisions and compared to proton-induced results at similar
incident energy. The comparison suggests excess direct photon
production in central $^{208}$Pb\/+\/$^{208}$Pb collisions
beyond that expected from proton-induced reactions.  
The result suggests 
modification of the prompt photon production
in nucleus-nucleus collisions, or additional contributions from
pre-equilibrium or thermal photon emission.
The result should
provide a stringent test for different reaction scenarios,
including those with quark gluon plasma formation, and may provide  
information on the initial temperature attained in these collisions.

We wish to express our gratitude to the CERN accelerator division for the
excellent performance of the SPS accelerator complex. We acknowledge with
appreciation the effort of all engineers, technicians and support staff who
have participated in the construction of this experiment. 
This work was supported jointly by
the German BMBF and DFG,
the U.S. DOE,
the Swedish NFR and FRN,
the Dutch Stichting FOM,
the Stiftung f{\"u}r Deutsch-Polnische Zusammenarbeit,
the Grant Agency of the Czech Republic under contract No. 202/95/0217,
the Department of Atomic Energy,
the Department of Science and Technology,
the Council of Scientific and Industrial Research and
the University Grants
Commission of the Government of India,
the Indo-FRG Exchange Program,
the PPE division of CERN,
the Swiss National Fund,
the INTAS under Contract INTAS-97-0158,
ORISE,
Grant-in-Aid for Scientific Research
(Specially Promoted Research \& International Scientific Research)
of the Ministry of Education, Science and Culture,
the University of Tsukuba Special Research Projects, and
the JSPS Research Fellowships for Young Scientists.
ORNL is managed by UT-Battelle, LLC, for the U.S. Department of Energy
under contract DE-AC05-00OR22725.
The MIT group has been supported by the US Dept. of Energy under the
cooperative agreement DE-FC02-94ER40818.

\begin{table*}[hbt]
\caption{ 
Various sources of systematical error in the 
WA98 158$~${\it A}~GeV $^{208}$Pb\/+\/$^{208}$Pb direct photon analysis
specified as a percentage of 
$(\gamma/\pi^{0})_{\rm Meas}$ (items a) ), $(\gamma/\pi^0)_{\rm Bkgd}$
(items b) ), or
$(\pi^{0})_{\rm Meas}/(\pi^0)_{\rm Bkgd}$ (item c) ).
The systematical errors are quoted at two $p_T$ values to give an 
indication of the dependence on transverse momentum. The errors
are estimated for the narrow shower identification criterion (S2).
The total estimated systematical error on 
$\gamma_{\rm Meas}/\gamma_{\rm Bkgd}$ is given as the quadratic sum
of the various contributions. See Ref.~[23] for 
full details.
}
\label{table3}
\begin{tabular}{lcccc}
Source of Error& \multispan2 
\hfill Peripheral Collisions $(20\%\ \sigma_{\rm mb})$ \hfill & 
\multispan2 \hfill Central\ Collisions\ $(10\%\ \sigma_{\rm mb})$ \hfill \\
& $p_T \approx 1.0$ GeV/c & $ p_T \approx 2.5$ GeV/c 
& $p_T \approx 1.0$ GeV/c & $ p_T \approx 2.5$ GeV/c \\
\hline
\ \ \ Charged Particle background$^+$ & 1.7 & 2.2 & 1.3 & 1.3 \\
\ \ \ $\gamma$ conversion correction$^+$ & 0.5 & 0.5 & 0.5 & 0.5 \\
\ \ \ Neutrons$^+$ & 0.6 & 1.0 & 0.9 & 1.9  \\
\ \ \ $\gamma$ reconstruction efficiency$^+$ & 2.0 & 2.0 & 2.0 & 2.0 \\
\hline
a) $\gamma$ yield measurement & 2.7 & 3.2 & 2.6 & 3.1 \\
\hline
\hline
\ \ \ $\gamma$ conversion correction$^*$ & 0.5 & 0.5 & 0.5 & 0.5 \\
\ \ \ $\pi^0$ yield extraction$^*$ & 0.3 & $<$0.1 & 5.1 & 1.0 \\
\ \ \ $\pi^0$ reconstruction efficiency$^*$ & 3.0 & 3.0 & 4.0 & 4.0 \\
\hline
a) $\pi^0$ yield measurement  & 3.1 & 3.0 & 6.5 & 4.2 \\
\hline
\hline
a) Non-target background & 1.5 & $<$0.1 & $<$0.1 & $<$0.1 \\
a) Energy scale calibration  & 0.9 & 1.7 & 0.8 & 1.7 \\
b) Detector acceptance & 0.5 & 0.5 & 0.5 & 0.5 \\
b) $\eta / \pi$ ratio, $m_T$-scaling & 2.9 & 3.2 & +3.4 (-4.8) & 
+3.7 (-5.2) \\
b) Other radiative decays & 1.0 & 1.0 & 1.0 & 1.0 \\
c) $\pi^0$ fit  & 1.6 & 6.8 & 2.9 & 0.4 \\
\hline
Total: (quadratic sum)  & 5.7 & 8.9 & +8.3 (-9.1) & +6.7 (-7.6) \\
\end{tabular}
$^+$ Included in $\gamma$ yield measurement error.
$^*$ Included in $\pi^0$ yield measurement error.
\end{table*}



\begin{references}
%
\bibitem{Abr97}
M.C.~Abreu et al., Phys. Lett. B {\bf 410}, 337 (1997).                        
%
\bibitem{And99}
E.~Andersen et al., Phys. Lett. B {\bf 449}, 401 (1999).
%
%
\bibitem{Fei76}
E.L.~Feinberg, Nuovo Cimento {\bf 34} A, 391 (1976).
%
\bibitem{Shu78} 
E.~Shuryak, Phys. Lett. B {\bf 78}, 150 (1978).
%
%
\bibitem{Kaj81}
K.~Kajantie and H.I.~Miettinen, Z. Phys. C  {\bf 9}, 341 (1981).
%
\bibitem{Hal82}
F.~Halzen and H.C.~Liu, Phys. Rev. D {\bf 25}, 1842 (1982).
%
\bibitem{Sin83}
B.~Sinha, Phys. Lett. B  {\bf 128}, 91 (1983).
%
\bibitem{McL85}
L.D.~McLerran and T.~Toimela, Phys. Rev. D {\bf 31}, 545 (1985).
%
\bibitem{Kap91}
J.~Kapusta, P.~Lichard, and D.~Seibert, Phys. Rev. D {\bf 44}, 2774 (1991).
%
%
\bibitem{Aur98}
P.~Aurenche, F.~Gelis, H.~Zaraket, and R.~Kobes, Phys. Rev. D {\bf 58}, 
085003 (1998).
%
\bibitem{Sri99}
D.K.~Srivastava, Eur. Phys. J. C {\bf 10}, 
487 (1999).
%
\bibitem{VW}    
W.~Vogelsang and M.R.~Whalley, J. Phys. G: Nucl. Part. Phys.  {\bf
  23}, A1 (1997).
%
%
%
\bibitem{Ake90}
HELIOS Collaboration, T.~{\AA}kesson et al., Z. Phys. C  {\bf 46}, 369 (1990).
%
\bibitem{Alb91}
WA80 Collaboration, R.~Albrecht et al., Z. Phys. C  {\bf 51}, 1 (1991).
%
\bibitem{Bau96}
CERES Collaboration, R.~Baur et al, Z. Phys. C {\bf 71}, 571 (1996).
%
\bibitem{Alb96}
WA80 Collaboration, R.~Albrecht et al., Phys. Rev. Lett. {\bf 76}, 
3506 (1996).
%
\bibitem{ref:sr94} 
D.K.~Srivastava and B.~Sinha, Phys.\ Rev.\ Lett.\ {\bf 73},
2421 (1994).
%
\bibitem{ref:ar95} 
N.~Arbex, U.~Ornik, M.~Pl\"umer, A.~Timmermann, and R.~M.~Weiner,
Phys. Lett. B {\bf 354}, 307 (1995).
%
\bibitem{ref:du95} 
A.~Dumitru, U.~Katscher, J.~A.~Maruhn, H.~St\"ocker, 
W.~Greiner, and D.~H.~Rischke, Phys. Rev. C {\bf 51}, 2166 (1995).
%
\bibitem{ref:ne95} 
J.~J.~Neumann, D.~Siebert, and G.~Fai, Phys. Rev. C {\bf 51},
1460 (1995).
%
\bibitem{ref:so97} 
J.~Sollfrank et al., Phys. Rev. C {\bf 55},
392 (1997).
%
\bibitem{misc:wa98:proposal:91}
WA98 Collaboration,
\newblock {\em Proposal for a large acceptance hadron and photon spectrometer},
  1991,
\newblock Preprint CERN/SPSLC 91-17, SPSLC/P260
%
\bibitem{wa98photonslong} 
WA98 Collaboration, M.M.~Aggarwal et al,  nucl-ex/0006007,
submitted to Phys. Rev. C.
%
%
\bibitem{Bor76} M.~Bourquin and J.-M.~Gaillard, Nucl. Phys. B {\bf
    114}, 334 (1976).
\bibitem{plb:ada95} E704 Collaboration, D.L.~Adams  et~al.,
  Phys. Lett. B {\bf 345}, 569 (1995).
%
\bibitem{prl:mcl83} E629 Collaboration, M.~McLaughlin et~al., 
Phys. Rev. Lett. {\bf 51}, 971 (1983).
%
\bibitem{zpc:bad86} NA3 Collaboration, J.~Badier et~al., 
Z. Phys. C {\bf 31}, 341 (1986).
%
\bibitem{rmp:owe87} J.F.~Owens, Rev. Mod. Phys. {\bf 59}, 465 (1987).
%
%
\end{references}
\end{document}